\documentclass[amssymb,pre,aps,twocolumn,showpacs,floatfix]{revtex4}

\usepackage{graphicx}

\begin{document}
\title{Early sedimentation and crossover kinetics\\
in an off-critical phase-separating liquid mixture}
\author{J. Colombani}
\email{Jean.Colombani@lpmcn.univ-lyon1.fr}
\author{J. Bert}
\affiliation{Laboratoire Physique de la Mati\`ere Condens\'ee et Nanostructures
(UMR CNRS 5586),
Universit\'e Claude Bernard Lyon 1, 6, rue Amp\`ere,
F-69622 Villeurbanne cedex, France}

\begin{abstract}
Early sedimentation in a liquid mixture off-critically quenched
in its miscibility gap
was investigated with a light attenuation technique.
The time evolution of the droplets distribution is characteristic of
an emulsion coalescing by gravitational collisions.
This sedimentation behaviour has given access to the phase-separating
kinetics and a crossover on the way toward equilibrium was observed,
which separates free growth from conserved order-parameter
coarsening with a crossover time fitting well to theoretical predictions.
\end{abstract}

\maketitle

\section{Introduction}

The study of liquid-liquid phase separation constitutes a unique
opportunity of observing the decay modes to equilibrium of a system
abruptly brought into a metastable or nonequilibrium state.
One benefits
here from the universal behaviour of the dynamical properties in the
vicinity of a consolute critical point.

The leading quantity
is the supersaturation $\Phi$, corresponding to the equilibrium volume fraction
 of the minority phase.
For values of $\Phi$ from $\frac{1}{2}$ to $0$
 ---or  equivalently from deep to shallow quenches
in the miscibility gap--- the following stages
 can be encountered~:

\begin{itemize}
\item For critical, i.e., quasisymmetric,
quenches ($\Phi\simeq\frac{1}{2}$), the phase-separating
mechanism is spinodal decomposition and has been abundantly
investigated.
Incipient domains from both phases grow from the most unstable wavelength
of the concentration fluctuations $\xi^-$ \cite{Cahn}.
Afterwards they coarsen by brownian collision-induced
coalescence and the time evolution of their mean size scales as
$t^\frac{1}{3}$ \cite{Wong}.
When they constitute a bicontinuous percolating medium, coarsening
continues through surface-tension driven mechanisms (Rayleigh-like
instability \cite{Siggia}, coalescence-induced coalescence \cite{Nikolayev})
which induce a coarsening law linear with time \cite{Wong}.

\item For off-critical, i.e., nonsymmetric, quenches,
the phase-separating regime is nucleation and
growth, where the nucleation has
recently been stated as always being heterogeneous.
This statement has been
inferred from the monodispersity \cite{Cumming} and the undercriticality
\cite{Buil} of the nucleated droplets size.
Growth proceeds through the
diffusion of one of the components from the supersaturated background to a
fixed number of growing nuclei ('free growth') with a
$t^ \frac{1}{2}$ law \cite{Langer}.
After an intermediate regime where
the volume fraction of minority phase reaches its equilibrium
value $\Phi$, two competing mechanisms are expected.
First when the solute-depleted layers around the growing droplets begin
to interact, the coarsening mechanism may become Ostwald ripening, i.e.,
evaporation-condensation (undercritical nuclei
dissipate into critical ones), obeying the $t^\frac{1}{3}$ 
Lifschitz-Slyozov (LS) law \cite{Akaiwa}.
Parallely, as for the spinodal decomposition,
when interfaces become sharp, coalescence may proceed through
brownian diffusion (BD) induced collisions of droplets, which implies also a
$t^\frac{1}{3}$ law, with a different prefactor from LS's one
\cite{Siggia}.
The latter process should prevail at high $\Phi$ values and late times.
\end{itemize}

The volume fraction of minority phase $\Phi_T$ separating these two scenarii
has been subject to
much discussion but the latest reliable value should be $\phi_T\simeq 30\%
$\cite{Perrot99}.
Few results
are available where $\Phi$ is systematically scanned between $0$ and $\phi_T
$ \cite{Perrot99,Hopkinson}.
Hence some points remain questionable among which
the precise domain of existence
of BD and LS coarsenings and the time of crossover between the free
diffusion ($t^\frac{1}{2}$) and conserved order-parameter ($t^\frac{1}{3}$)
regimes.

Whatever the regime, when the droplet size reaches a threshold value,
gravity begins to prevail and sedimentation occurs.
In the critical regime, two contributions to the study of sedimentation
have to be mentionned.
The first one identifies the successive stages during the prevailing of
gravity \cite{Chan}: Macroscopic convection, sedimentation of droplets
leading to a $t^3$ growth law, appearing of a meniscus sharing the mixture 
in two macroscopic phases, residual sedimentation.
The second one identifies
the growth law during this residual sedimentation stage as
$t^{0.27}$, as a consequence of coalescence by both sedimentation
and brownian diffusion \cite{Cau}.
Besides, sedimentation in the off-critical regime has never been
systematically investigated and it is our aim to explore it.

Unexpectedly, this sedimentation study gives the opportunity to probe early
times of the separation dynamics, delicate to access with
light scattering methods for instance \cite{Cumming}.
Thereby we present here experimental evidence of a crossover on the
way toward equilibrium of a phase-separating mixture.

\begin{figure}
\includegraphics[width=\linewidth]{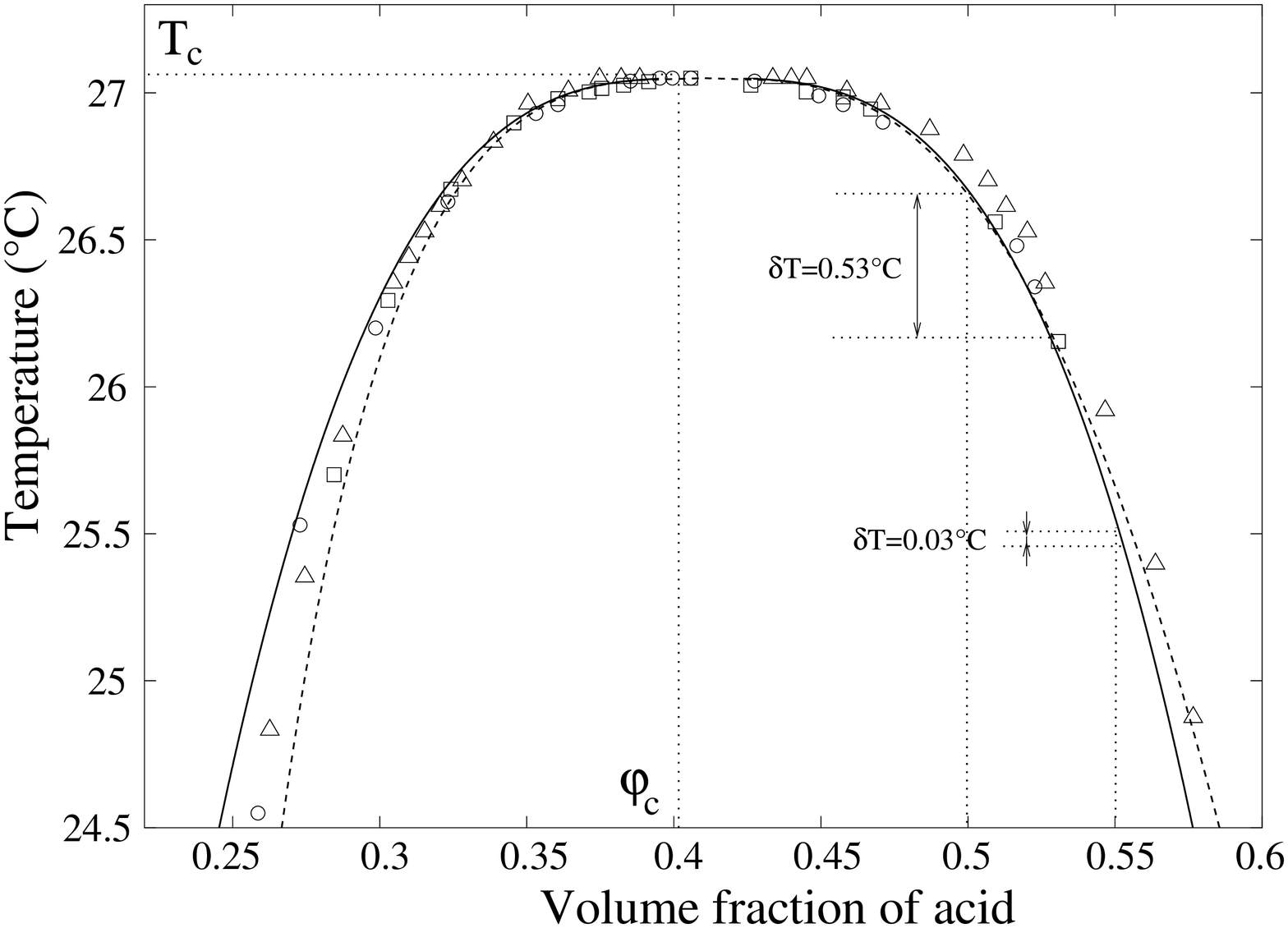}
\caption{Phase diagram ($\phi_{acid},T$) of the isobutyric acid-water mixture
from Woermann \textit{et al.} \cite{Woermann} ($\bigcirc$), Chu \textit{et al.}
\cite{Chu} ($\square$), Zhuang \textit{et al.} \cite{Zhuang} ($\triangle$),
Krall \textit{et al.} \cite{Krall} (dashed line) and Andrew \textit{et al.}
\cite{Andrew} (solid line).
The critical temperatures have been matched to ours : $T_c$=27.05$^\circ$C.
Our shallowest and deepest quenches are also represented.}
\label{diagram}
\end{figure}

\section{Experiments}

For this purpose, we have chosen the water-isobutyric acid
mixture, taking advantage from its room-temperature miscibility gap
and from the complete knowledge of its physicochemical properties.

Numerous experimental determinations of the phase diagram of
this system may be found in the literature.
To get a clear view of the experimental uncertainty
(mentionned by Baumberger \textit{et al.} \cite{Baumberger})
on the coexistence curve,
which is of former importance for the computation
of the volume fraction of the growing phase $\Phi$,
we gathered the most representative experimental phase diagrams
on a single volume fraction-temperature plot (cf. Fig. \ref{diagram}).
The new processing of the values of Hamano \textit{et al.}
\cite{Hamano} by Krall \textit{et al.} \cite{Krall} and the more recent
interpretation of Greer's density measurements \cite{Greer} by Andrew
\textit{et al.} \cite{Andrew} have been chosen rather than the original
corresponding works.
The acid mass fraction values $c_{acid}$
of Krall \textit{et al.} \cite{Krall}, Woermann \textit{et al.} \cite{Woermann},
Zhuang \textit{et al.} \cite{Zhuang} and Chu \textit{et al.} \cite{Chu}
have been turned into volume fraction values $\phi_{acid}$ through
$\phi_{acid}=(\rho_{phase}/\rho_{acid})c_{acid}$, the densities $\rho$
of the two phases and of isobutyric acid being taken from \cite{Greer}.
The critical temperature $T_c$ shows a dispersion of more
than one degree among the experiments, most certainly due to ionic impurities,
which is of little consequence on the critical behaviour \cite{Greer,Cohn}.
So all curves have been adjusted to our experimental value
$T_c=27.05^\circ C$.
The data of Andrew \textit{et al.} are quite recent and situated
in the mean range of the other results \cite{Andrew}.
Accordingly their expression for the miscibility gap
$\Delta\phi_{acid}=\Delta\phi_0 \epsilon^\beta$
with $\Delta\phi_{acid}=\phi_{acid}-\phi_c$, $\phi_c=0.4028$,
$\Delta\phi_0=1.565$, $\beta=0.326$ and $\epsilon=(T_c-T)/T_c$
the reduced temperature, has been chosen for our computation of $\Phi$.

For the correlation length of the concentration fluctuations along
the binodal line, the expression $\xi^-=\xi_0\epsilon^{-\nu_\xi}$ with
$\xi_0=1.8 \AA\ $
and $\nu_\xi=0.63$ has been chosen \cite{Krall}.
The viscosities are $\eta=(\eta^B+A'\epsilon^\frac{1}{3})\epsilon^{-0.04}$
for the acid-rich phase and
$\eta'=(\eta^B-A'\epsilon^\frac{1}{3})\epsilon^{-0.04}$ for the water-rich phase,
with $\eta^B=1.89$ mPl and $A'=2.60$ mPl \cite{Krall}.
The diffusion coefficient of the mixture along the acid-rich branch of
the coexistence curve is deduced from the above quantities
thanks to a Stokes-Einstein relation: $D^-=k_B T/(6\pi\eta\xi^-)$.

\begin{figure}
\includegraphics[angle=-90,width=\linewidth]{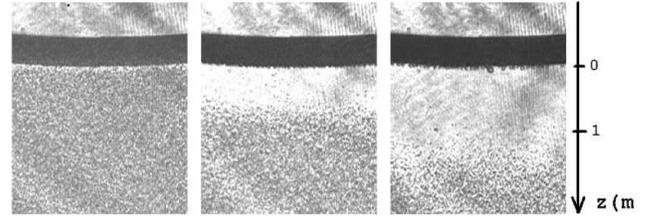}
\caption{Photographs of the sedimenting droplets in a mixture with 
$\phi_{acid}$=54\% for $\delta T$=0.13 K, 3.5, 23.5 and 43.5 min.
after the quench.}
\label{photo}
\end{figure}

To perform the quenches, an optical fused quartz cell containing the mixture
is inserted inside a hollow copper block where
a thermostat ensures a water circulation,
providing a temperature stabilization of the system within 0.01 K.
The cell (optical pathlength 0.1 cm, 1 cm
wide and 3 cm high) is illuminated by a laser beam
and is observed by means of a Charge-Coupled Device camera.
For each studied concentration, the temperature of the coexistence curve has
 been visually determined (cloud point method) by a slow decrease of the
temperature from the one-phase mixture (0.01 K temperature steps, each
 followed by a 20 minutes stabilization).

Before each run, an energetic stirring is performed followed by a
12 hours annealing in the
one-phase region 0.05 K above the coexistence curve. Then, the
mixture is rapidly quenched through the binodal line
($\delta T=$0.03 to 0.53 K below it) and this
incursion in the miscibility gap leads to the phase separation
(cf. Fig. \ref{photo}).
In each case, the volume fraction of acid $\phi_{acid}$
(50 to 55\%) and the reduced temperature $\epsilon_f=(T_c-T_f)/T_c$
(2.3 to 6.4$\times 10^{-3}$ with $T_f$ the absolute final temperature)
are chosen in order to
induce a volume fraction $\Phi$ of the growing phase between
0.3 and 10.5\%.
The water-rich phase (density $\rho'\simeq 998$\ kg m$^{-3}$)
has been chosen as the nucleating and sedimenting phase to prevent
wetting effects.
Indeed the isobutyric acid-rich phase (density
$\rho\simeq988$\ kg m$^{-3}$) is known to preferentially wet the cell walls
\cite{Andrew}.

\begin{figure}
\includegraphics[width=\linewidth]{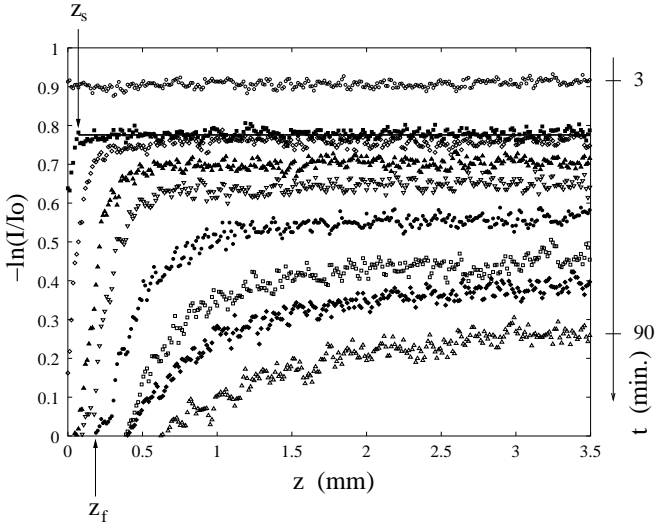}
\caption{Logarithm of the light attenuation $-ln(I/I_0)$ across the cell
as a function of the vertical
position $z$ and elapsed time $t$ for a volume fraction of acid
$\phi_{acid}=54 \%$ and a quench depth $\delta T=$0.37 K.}
\label{attenuation}
\end{figure}

The evolution of the transmitted intensity along the vertical axis of the cell
$I(z)$ is extracted from the video output
---kept in its linear response domain---
via an image processing software (cf. Fig. \ref{photo})
An averaging of $I(z)$ along an horizontal segment in the middle of the cell
is carried out for each value of $z$ to gain noise reduction.
Then the light attenuation $I(z)/I_0$ is computed for several times
during the phase separation.
$I_0$ stands for the unscattered light intensity refracted by the homogeneous
(droplet-free) mixture.
This light attenuation is mainly due to scattering.
The refraction indices of the two phases are very similar ($\Delta n\sim 10^{-2}$
in our $\epsilon$ range \cite{Andrew}), which is likely
to induce a low multiscattering.
So in our $\Phi$ range and assuming a small polydispersity,
the light attenuation $I(z)/I_0$ is linked at a first order to the
concentration of droplets $n(z)$ through the Lambert-Beer law
$n(z)\sim -ln(I(z)/I_0)$.
Therefore the evolution of the $-ln(I(z)/I_0)$ curves with time reports on
the change of spatial distribution of droplets during sedimentation as
displayed in Fig. \ref{attenuation}.

The droplets being larger than the laser wavelength $\lambda=$0.532 $\mu$m
(see below), the attenuation
should also be inversely proportional to their squared radius,
which could have a slight influence on $-ln(I/I_0)$.
Therefore this expression constitutes a qualitative evaluation of $n$
but our analysis does not
require a more precise determination, dealing only with abrupt slope changes
in the behaviour of $n$.

The time origin has been chosen as the cloud appearance time and the
space origin as the lower limit of the liquid-air meniscus
(cf. Fig. \ref{photo}).

Right after the quench, the cell becomes uniformly filled
with erratically moving droplets
forming an opalescent mist (horizontal upper line in Fig. \ref{attenuation}).
Then in the vicinity of the meniscus the droplet concentration
progressively decreases down to a zero value
owing to an overall vertical motion of the droplets only visible
at the top (clarification zone, droplet-free)
and the bottom (sedimentation
layer, not shown in Fig. \ref{photo} and \ref{attenuation}) of the vessel.
Afterwards, the initial point $z_f$ of the curves moves down
and the value of the plateau sinks.
The first feature reflects a growth of the clarification zone and the second one
owes to the decrease of the bulk droplet concentration due to coalescence
yielded by brownian and/or gravitational collisions.

\section{Coarsening mechanism during sedimentation}

In order to highlight the respective role of the brownian and
gravitational processes, the 
knowledge of the P\'eclet number $Pe=v R/D$ is needed
($v$ mean sedimentation velocity,
$R$ radius and $D$ diffusion coefficient of the droplets).
Indeed this number compares the mean sedimentation velocity $v$
to a brownian velocity $D/R$.

With this purpose, we turn to the clarification zone behavior.
Its growth can be followed in Fig. \ref{attenuation} through the increase
with time of the interceipt $z_f$ with time of the light attenuation
curve with the abscissa axis.
In other words, $z_f$ corresponds to the upper limit of the sedimentation front.
At this point, the droplets are very scarce
so they can be considered as evolving in a
quasi-infinitely dilute regime.
Therefore the sedimentation-induced convective part of their motion
is insignificant \cite{Davis85} and their velocity $v_f$ corresponds to
the stationary velocity of isolated droplets of radius $R_f$, given by
the Hadamard formula \cite{Hadamard}:
\begin{equation}
v_f=\frac{2(\eta'+\eta)(\rho'-\rho)R_f^2g}{3(3\eta'+2\eta)\eta}
\label{Hadamard}
\label{rayon}
\end{equation}
with $g$ the gravitational acceleration, $\eta'$ and $\eta$ respectively the
viscosity of the water-rich drop and of the acid-rich surrounding fluid,
$\rho'$ and $\rho$ their respective density.
To assess the validity of this expression, two checkings have been carried out.
This formula is valid provided that inertia effects are negligible.
This requirement is guaranteed by a low Reynolds number
$Re=R_f v_f \rho/\eta$, always smaller than 2.10$^{-6}$ in our case.
Secondly, even if long-range interdroplet
hydrodynamic interactions induced an average settling velocity $v_{av}$
lower than the Hadamard velocity $v_f$,
the hindered settling function $f(c)=v_{av}/v_f$ would take values between
$\frac{1}{2}$ and 1 for $\Phi$ ranging from 1 to 10\% \cite{Davis85}.
So these interactions would not yield a noticeable change of
the value of the settling velocity.

\begin{figure}
\includegraphics[width=\linewidth]{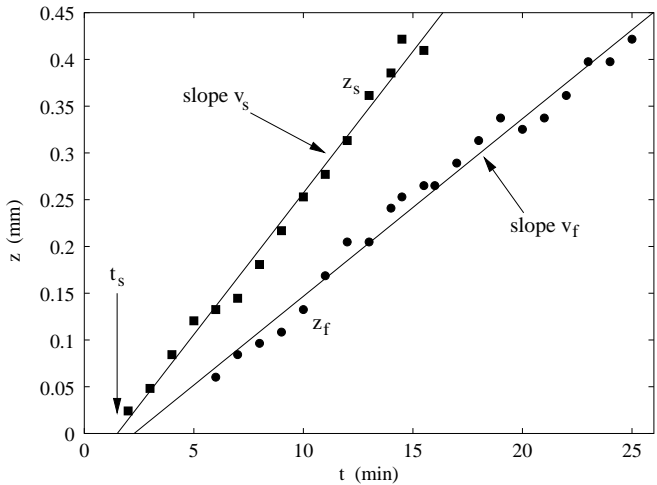}
\caption{Position of the beginning point $z_f$ and of the start
of the plateau $z_s$ of the light attenuation curves as a function of time
in the ($\phi_{acid}=52\%,\delta T=$0.12 K) case.
The lines are least-square fits of the values.
The $t_s$ time is also shown.}
\label{position}
\end{figure}

As $v_f$ is given by the slope of the $z_f(t)$ line (cf. Fig. \ref{position}),
the droplet radius $R_f$ at the appearance of the clarification zone can be
computed from Eq. \ref{rayon}.
$R_f$ ranges between 3.3 and 8.9 $\mu$m.
Knowing that the diffusivity of spherical droplets immersed
in a liquid of viscosity $\eta$
is $D_f=k_B T/(5\pi\eta R_f)$
\footnote{This expression is obtained in writing, in
the Stokes-Einstein expression of the diffusivity of one droplet
$D=\frac{k_BT}{\mu}$, the mobility $\mu$ as
$v_f/F_g$ with $v_f$ the Hadamard sedimentation velocity of Eq. \ref{Hadamard}
taken with $\eta=\eta'$ and $F_g=(\rho'-\rho)g\frac{4}{3}\pi R_f^3$ the
gravitational force exerting on the droplet.},
their P\'eclet number $Pe_f=v_f R_f/D_f$ can be computed:
\begin{equation}
Pe_{f}=\frac{15\pi\eta^2(3\eta'+2\eta)v_f^2}{2k_BT(\eta'+\eta)(\rho'-\rho)g}.
\end{equation}
$Pe_f$ is found to range between 8 ($\phi_{acid}=50$\% and $\delta T=0.53
^{\circ}$C) and 371 ($\phi_{acid}=53$\% and $\delta T=0.13^{\circ}$C).
Recalling that these droplets experience almost no coalescence owing to their
dilute environment, their growth is the slowest of the vessel,
their radius is the smallest, and their
velocity is minimal among all droplets,
so we can argue that $Pe\gg 1$ in the bulk sedimenting liquid.
Therefore gravitation-induced hydrodynamic interactions
should prevail over brownian diffusion as the coalescence mechanism.

To confirm this statement, we consider 
theoretical studies on
coalescence of non-brownian sedimenting polydisperse droplets immersed
in an immiscible fluid \cite{Wang95}.
The evolution of the volume fraction of droplets with position in the vessel
at different times (Fig. 7 of Ref. \cite{Wang95})
shows a strong closeness with our light attenuation curves of Fig.
\ref{attenuation}.
Therefore the droplet density in our sedimenting mixture displays
the same evolution as predicted for non-brownian droplets
and, at variance with some predictions \cite{Baumberger,Cau},
the driving force of coalescence is likely to consist only
of gravitational collisions.

\section{Phase separation dynamics}

What can we learn now about the drops growth law during this stage
of phase separation?
To address this point,
we focus on the early appearance of sedimentation.
When growing droplets experience a gravitational force overwhelming brownian
diffusion, the symmetry of their displacement is broken
and they exhibit an average descending motion.
This incipient settling motion can be traced in Fig. \ref{attenuation}
through the displacement with time of the beginning $z_s$ of the curves
plateau.
In other words, we follow with $z_s$ the initial mean trajectory of
the droplets settling from the meniscus.
We concentrate on early times, where the interceipt of the rising curve
and the plateau is unambiguous.
If the P\'eclet number $Pe_s$, mean sedimentation velocity $v_s$, and
diffusivity $D_s$
of these settling droplets are known independantly, their radius 
can be computed, considering the definition of $Pe_s$,
via $R_s=\sqrt{Pe_s D_s/v_s}$.
Using the above-mentionned Stokes-Einstein expression for $D_s$,
one gets $R_s=\sqrt{Pe_s k_B T/(5\pi\eta v_s)}$.

The value of $\eta$ is given above and $v_s$ is accessible through
the slope of the $z_s(t)$ line (cf. Fig. \ref{position}).
So at this point, we need a determination of the only missing value,
$Pe_s$, to be able
to calculate $R_s$ at least at one particular time of the coarsening.

The sedimentation regime is entered when the P\'eclet number sufficiently
exceeds unity.
We will tentatively consider that this is the case
when the sedimentation velocity becomes one order of magnitude larger
than the diffusive velocity.
The time $t_s$ of this start of sedimentation is measured in extrapolating
the $z_s(t)$ curve to zero.
$z_s(t)$ has been seen to remain linear at early times,
which enables a well-defined extrapolation of $t_s$ (cf. Fig. \ref{position}).
So at $t_s$ we assume that $Pe_s=10$ and $R_s$ can be computed with
the above-mentioned formula.
As expected for droplets at the non sedimenting-sedimenting transition,
$R_s$ ranges between 1.0 and 3.2 $\mu$m for our quenches.

Therefore we use our knowledge of the sedimentation behaviour as a probe
to determine the droplet radius $R_s$ just when sedimentation sets in.
Accordingly, the $R_s$ values of the different
quenches are characteristic of the early coarsening mechanisms preceding
sedimentation (free growth, BD, LS) and not related to the settling
behaviour itself.

Concerning the choice of $Pe_s$,
Wang and Davis have performed computations in an immiscible mixture
experiencing simultaneous brownian and gravitational collisions
\cite{Wang96}.
Using their predictions, we have computed the mean radius $R_s$ of the
coalescing droplets after a time estimated as $t_s$ 
in the most unfavourable
case of viscosity, dispersity and settling velocity.
We find that $R_s\simeq$ 1.5 $R_0$ ($R_0$ initial radius) for an initial
P\'eclet number $Pe_s=1$ and $R_s\simeq$ 2.4 $R_0$ for $Pe_s=10$.
Therefore we can conclude that 1) whatever the choice of $Pe_s$,
gravity has not
significantly modified the coarsening dynamics after an elapsed time $t_s$ and 2)
this minor influence on $R_s$, if any, is comparable for $Pe_s$ ranging
from 1 to 10.
Furthermore $R_s\sim \sqrt{Pe_s}$ which induces a weak influence of the
choice of $Pe_s$, between 1 and 10, on the computation of $R_s$.

\begin{figure}
\includegraphics[width=\linewidth]{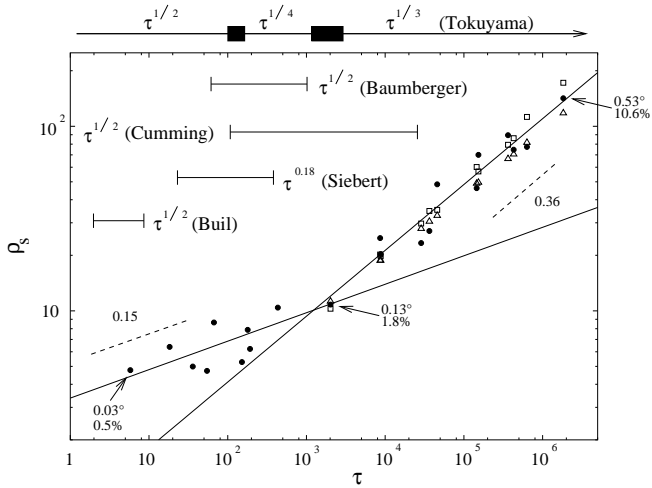}
\caption{Evolution with reduced time $\tau$ of the reduced droplet
size $\rho_s$ at the end of the non-sedimenting regime (black dots).
The open triangles and squares are respectively the theoretical values of
$\rho_s$ for an Ostwald ripening and a brownian diffusion-coalescence
growth.
It should be noticed that each point corresponds to one experiment.
Ranges of growth exponents available in the literature for $\Phi<10$\%
have also been added (references given in the text).
The quench depth $\delta T$ and the volume fraction of the growing phase
$\Phi$ have been mentionned for some representative points.}
\label{sizeS}
\end{figure}

To allow comparison between the dynamics of all experiments, we put the time
and radius values in a dimensionless form : $\tau=t/t_c$ and
$\rho=R/R_c$.
The renormalization quantities are the radius of an initial critical nucleus
$R_c=\alpha/\Phi$ and the relaxation time of this nucleus
$t_c=D^-\alpha^2/\Phi^3$,
with $\alpha$ a capillary length estimated as $\alpha=\xi^-/3$
\cite{Langer}.

We benefit now from one point $(\tau_s,\rho_s)$
for each experiment, covering five decades of reduced time.
Fig. \ref{sizeS} displays the evolution of $\rho_s$ as a function of $\tau_s$.
We first ascertain that all experiments merge onto a single curve, whatever the
initial supersaturation.
The second striking feature lies in the presence of a crossover between two
coarsening stages.
The first behaviour can tentatively be fitted by a power law
of exponent $\simeq 0.15$.
Such a slow radius evolution is expected at the end of the free growth regime
and reflects the overlapping of the solute-depleted shells around the nuclei
\cite{Tokuyama}.
The second behaviour displays a growth exponent $0.36\simeq\frac{1}{3}$ which is
the well-known exponent of interface-reduction coarsenings (LS and BD).
To our knowledge this constitutes the first experimental observation of
the crossover
between free growth and constant volume-fraction coarsening in a liquid system.
This crossover takes place at $\tau_ {CO}\simeq 1\times 10^3$.

The experimental and theoretical growth laws available for $\Phi<10$\%
in the literature have also been added in Fig. \ref{sizeS}.
Surprisingly, no other measurements exist in the $t^{\frac{1}{3}}$ range
for $\Phi<10\%
$ except those of Wong \& Knobler which seem to display a
progressive change of slope and are therefore delicate to deal with \cite{Wong}
and those of White \& Wiltzius, the dimensionless parameters of which
are not available \cite{White95}.
Concerning the end of free diffusion, our growth exponent is equivalent
to Siebert \& Knobler's one (0.18) \cite{Siebert}
and is compatible with the theoretical
prediction of Tokuyama \& Enomoto ($\frac{1}{4}$) \cite{Tokuyama}.
The validity domain of this intermediate stage is also compatible
with the $t^{\frac{1}{2}}$ range  of Buil \textit{et al.} \cite{Buil}
but is not consistent with those of Baumberger \textit{et al.}
\cite{Baumberger} and Cumming \textit{et al.} \cite{Cumming}.
The latter dealt with polymer blends and this could account for the
dynamics discrepancy or at least for the difficulty of computing equivalent
dimensionless quantities in simple (acid/water) and complex (polymer blends)
fluids.
The inconsistency with the former remains unexplained.
The crossover time
compares well with Tokuyama \& Enomoto's theoretical value
$1.1\times 10^3<\tau_{CO}^{theo}<2.8\times 10^3$,
the lower value computed at $\Phi=1\%$ and the upper one at $\Phi=10\%
$ \cite{Tokuyama}.

As the exact mechanism of the late decompositional stage still remains
debatable \cite{White95}, we have drawn in Fig. \ref{sizeS} the theoretical
values of $\rho_s(\tau)$ predicted by the Lifschitz-Slyozov and 
brownian diffusion coarsening laws, computed
without adjustable parameters.
Written in reduced units, the expressions are
$\rho_s=[1+(4/9)f(\Phi)\tau]^{\frac{1}{3}}$ ($f(\Phi)$ being a
numerically estimated function)
for evaporation-condensation \cite{Akaiwa} and
$\rho_s=[144\Phi/(\ln(0.55R_s/\delta))]^\frac{1}{3}$ (with $\delta\simeq\xi^-$)
for brownian diffusion coalescence \cite{Siggia}.
Unfortunately, though experimental and theoretical values superimpose 
\footnote{This superimposition justifies
\textit{a posteriori} our $Pe_s=10$ choice for the computation of $\rho_s$.
Nevertheless a change in $Pe_s$ would have only implied a vertical translation of
the values but would have had no influence on the kinetics and crossover time.},
the dispersion of our values is too large to discriminate between
the LS and BD laws, and their closeness banish any hope to do so
from $\rho(\tau)$ curves.

\section{Conclusion}

We have determined the leading coarsening mechanism
(gravitational collisions) during the sedimentation regime
in a homogeneous mixture plunged in a metastable state.
Besides, thanks to the superposition of numerous sedimentation
experiments in such systems, the
universal crossover time between free growth and ripening by diffusion of a
conserved order-parameter has been measured and agrees with theoretical
predictions.
Unfortunately the nature of the second mechanism has not been settled
and the existence range of Ostwald ripening and brownian coalescence coarsening
remains to be established.
Finally, an experimental access to the complete scenario of the off-critical
coarsening from the $t^\frac{1}{2}$ to the $t^\frac{1}{3}$ law would bring
a comprehensive view of the phase-separation process and a complete
verification of the nucleation and growth theory.

\begin{acknowledgments}
We acknowledge CNES (french space agency) for financial support,
B\'en\'edicte Herv\'e, Laurence Heinrich and Richard Cohen
for experimental help, and Jean-Pierre Delville, R\'egis Wunenburger,
Christophe Ybert and Elisabeth Charlaix for fruitful discussions.
\end{acknowledgments}

\end{document}